\documentclass[a4paper, conference]{IEEEtran}
\IEEEoverridecommandlockouts

\usepackage[acronyms]{glossaries}
\newacronym{3gpp}{3GPP}{3rd Generation Partnership Project}
\newacronym{5g}{5G}{fifth generation}
\newacronym{ai}{AI}{artificial intelligence}
\newacronym{awgn}{AWGN}{additive white gaussian noise}
\newacronym[\glslongpluralkey={angles of arrival}]{aoa}{AoA}{angle of arrival}
\newacronym[\glslongpluralkey={angles of departure}]{aod}{AoD}{angle of departure}
\newacronym{bim}{BIM}{beam index map}
\newacronym{bs}{BS}{base station}
\newacronym[\glslongpluralkey={channel covariance matrices}]{ccm}{CCM}{channel covariance matrix}
\newacronym{ce}{CE}{channel estimation}
\newacronym[\glslongpluralkey={channel state information}]{csi}{CSI}{channel state information}
\newacronym{cme}{CME}{conditional mean estimator}
\newacronym{cg}{CG}{conditionally Gaussian}
\newacronym{ckm}{CKM}{channel knowledge map}
\newacronym{cnn}{CNN}{convolutional neural network}
\newacronym{dft}{DFT}{discrete Fourier transform}
\newacronym{dl}{DL}{deep learning}
\newacronym{dnn}{DNN}{deep neural network}
\newacronym{elbo}{ELBO}{evidence lower bound}
\newacronym{em}{EM}{expectation maximization}
\newacronym{evd}{EVD}{eigendecomposition}
\newacronym{gan}{GAN}{generative adversarial network}
\newacronym{gcs}{GCS}{global coordinate system}
\newacronym{gmm}{GMM}{Gaussian mixture model}
\newacronym{gnss}{GNSS}{global navigation satellite system}
\newacronym{lmmse}{LMMSE}{linear minimum mean square error}
\newacronym{los}{LOS}{line of sight}
\newacronym{ls}{LS}{least squares}
\newacronym{mf}{MF}{matched filter}
\newacronym{mimo}{MIMO}{multiple-input multiple-output}
\newacronym{ummimo}{umMIMO}{ultra-massive multiple-input multiple-output}
\newacronym{ml}{ML}{machine learning}
\newacronym{mmse}{MMSE}{minimum mean square error}
\newacronym{mnse}{MnSE}{mean normalized spectral efficiency}
\newacronym{mse}{MSE}{mean square error}
\newacronym{mt}{MT}{mobile terminal}
\newacronym{nn}{NN}{neural network}
\newacronym{nlos}{NLOS}{non-line of sight}
\newacronym{nr}{NR}{new radio}
\newacronym{nmse}{NMSE}{normalized mean square error}
\newacronym{ofdm}{ODFM}{orthogonal frequency-division multiplexing}
\newacronym{pc}{PC}{pilot contamination}
\newacronym{pdf}{PDF}{probability density function}
\newacronym{pg}{PG}{path-gain}
\newacronym{quadriga}{QuaDRiGa}{QUAsi Deterministic RadIo channel GenerAtor}
\newacronym{rss}{RSS}{received signal strength}
\newacronym{relu}{ReLU}{rectified linear unit}
\newacronym{rf}{RF}{radio frequency}
\newacronym{rx}{rx}{receiver}
\newacronym{simo}{SIMO}{single-input multiple-output}
\newacronym{sinr}{SINR}{signal to interference plus noise ratio}
\newacronym{snr}{SNR}{signal-to-noise ratio}
\newacronym{se}{SE}{spectral efficiency}
\newacronym{sow}{SOW}{statement of work}
\newacronym{tdd}{TDD}{time division duplex}
\newacronym{tx}{tx}{transmitter}
\newacronym{ue}{UE}{user equipment}
\newacronym{uma}{UMa}{Urban Macro-Cell}
\newacronym{ula}{ULA}{uniform linear array}
\newacronym{vae}{VAE}{variational autoencoder}
\newacronym{vi}{VI}{variational inference}
\newacronym{wlan}{WLAN}{wireless local area network}
\newacronym{wss}{WSS}{wide sense stationary}
\usepackage{siunitx}
\usepackage{amsmath}
\usepackage{bm}
\usepackage{amssymb}
\usepackage{cleveref}
\usepackage{tikz}
\usepackage{pgfplots}
\usepackage{pgfplotstable}
\usetikzlibrary{patterns}
\usepackage{subfig}
\usepackage{cite}
\usepackage{amsmath,amssymb,amsfonts}
\usepackage{algorithmic}
\usepackage{graphicx}
\usepackage{textcomp}
\usepackage{xcolor}
\def\BibTeX{{\rm B\kern-.05em{\sc i\kern-.025em b}\kern-.08em
    T\kern-.1667em\lower.7ex\hbox{E}\kern-.125emX}}
\begin{document}

\title{Lightweight Beam Index Map Using Coupled Gaussian Mixture Models}

\author{\IEEEauthorblockN{Amar Kasibovic, Franz Weißer, Wolfgang Utschick}
\IEEEauthorblockA{\textit{TUM School of Computation, Information and Technology, Technical University of Munich, Germany} \\
\{amar.kasibovic, franz.weisser, utschick\}@tum.de}
}

\maketitle

\begin{figure}[b]
    \onecolumn
    \scriptsize
    © 2026 IEEE.  Personal use of this material is permitted.  Permission from IEEE must be obtained for all other uses, in any current or future media, including reprinting/republishing this material for advertising or promotional purposes, creating new collective works, for resale or redistribution to servers or lists, or reuse of any copyrighted component of this work in other works.
    \vspace{-2.5cm}
    \twocolumn
\end{figure}

\begin{abstract}
    This paper addresses the beam alignment problem in \acrshort{mimo} systems from a decentralized, \acrfull{mt}-centric perspective. 
    We propose a lightweight machine learning approach that leverages position information to perform beam selection without relying on exhaustive search or strong base station coordination.
    Specifically, we model the joint distribution of \acrshort{mt} positions and channel observations using a coupled \acrfull{gmm}, enabling the construction of a \acrfull{bim} that directly associates spatial locations with codebook entries.
    To account for practical hardware constraints, we introduce a refinement procedure that adapts the learned statistical model to fixed codebooks.
    The resulting method is computationally efficient and suitable for deployment on resource-constrained devices.
    Simulation results on the DeepMIMO and \acrshort{quadriga} datasets demonstrate that the proposed approach outperforms clustering-based fingerprinting methods and achieves competitive performance compared to exhaustive search, while significantly reducing complexity and overhead.
\end{abstract}

\section{Introduction}
    The next generation of wireless communication systems is expected to increasingly rely on \gls{ai}-based technologies to achieve higher performance under much more stringent constraints~\cite{VisionOf6GWirelessSystems-saad2020}.
    By exploiting massive amounts of data, \gls{ml} models can gather prior information to improve the performance of channel coding, \gls{mimo} precoding, \gls{ofdm} modulation, channel estimation, and other physical‑layer tasks~\cite{6gWirelessNetworks-zhang2019}.
    In this paper, we focus on the beam alignment problem---a subproblem of the \gls{mimo} precoding problems---which consists of selecting the optimal beam pair from predefined codebooks at the \gls{bs} and the \gls{mt}. 
    Specifically, the \gls{tx} applies precoding, while the \gls{rx} applies combining, and the goal is to maximize the \gls{snr} of the received signal~\cite{mmWaveBeamforming-Hur2013}.
    Beam alignment is particularly relevant at higher frequencies and with larger antenna arrays~\cite{TerahertzCommunicationsFor6gSurvey-Jiang2024}.

    The classical approaches to beam alignment rely on beam training or estimation.
    The former includes exhaustive search algorithms standardized by the IEEE for \gls{wlan} systems~\cite{IEEE802.11ad-2012, IEEE802.11ay-2021} and are part of the \gls{3gpp} discussions for the next release of \gls{5g} \gls{nr} systems~\cite{3gpp.38.843rel19}.
    Exhaustive approaches measure the \gls{snr} for each possible beam pair to perform the selection.
    The estimation approaches, instead, commonly determine underlying physical propagation parameters, such as the propagation angles of electromagnetic waves, for later use in the beam generation process \cite{beamTracking-Va2016, robustBeamTracking-Jayaprakasam2017}.
    Some works improved the beam estimation by incorporating \gls{ai}-based methodologies, for example, by training \gls{ml} models using multi-modal data, which includes, together with \gls{csi}, additional information such as \glspl{mt}' positions, scatterer positions, or data provided by a camera or a Radar sensor~\cite{multimodalDeepLearninMMWaveBeamPrediction-Shi2024}.
    Such works demonstrated that modeling the site-specific characteristics of a propagation environment can benefit the communication system.
    
    In this work, we focus on methods that define a relation between the \glspl{mt}' positions and channel modeling parameters that are connected to the beam selection task.
    According to the definition presented in~\cite{environmentAwareChannelKnowledgeMap-Zeng2024}, methods that establish a relationship between position and channel parameters are called \glspl{ckm}.
    In particular, methods that provide a beam index for a given position are called \glspl{bim}.
    We can consider the works in~\cite{environmentAwareBeamAlignment-Wu2021, probabilisticPositionAidedBeamSelection-Chege2025} as examples of \glspl{bim}.
    The authors of~\cite{environmentAwareBeamAlignment-Wu2021}, for example, construct a \gls{bim} by using the K-nearest neighbors algorithm.
    The authors of~\cite{probabilisticPositionAidedBeamSelection-Chege2025}, on the other hand, construct a probabilistic model to find the relation between the position information and the selected beam by modeling a joint probability mass function of the two quantities.

    \textit{Contributions}: In this work, we shift from the centralized \gls{bs} perspective to the decentralized network-edge perspective of \glspl{mt}.
    We rely on the position information available to the multi-antenna \glspl{mt} to perform the combining task in the case of downlink transmission. 
    The proposed method can be applied at the \gls{mt} with minimal dependence on the \gls{bs} operation, potentially reducing its coupling to standardized beam alignment procedures.
    We exploit a generative model based on the coupled \gls{gmm} presented in~\cite{noPilotsNoProblem-Weisser2025} to define a \gls{bim} that allows selecting a codebook entry at each possible location.
    We show that the proposed method operates using an extremely lightweight \gls{ml} model deployed on the \gls{mt}, with a reduced number of parameters.
    We propose a refinement of the map provided by the coupled \gls{gmm} to better adapt it to any given fixed codebook.
    Finally, we demonstrate that our method outperforms both exhaustive search and fingerprinting methods on two datasets.
    The simulation code is publicly available.\footnote{https://github.com/amark999/lightweight-beam-index-mapping}
\section{System Model}  \label{sec:system-model}

We consider a downlink, single-user, \gls{mimo} system, under static and narrowband conditions, where a single \gls{bs} is used and is equipped with \gls{ula} of \( N_{\text{tx}} \) antennas and a \gls{mt} is equipped with an \gls{ula} of \( N_{\text{rx}} \) antennas.
The wireless channel is thus described by the matrix \( \bm{H} \in \mathbb{C}^{N_{\text{rx}} \times N_{\text{tx}}} \).

In a single-stream downlink transmission, the \gls{bs} transmits a symbol \( s \in \mathbb{C} \) after precoding it via the vector \( \bm{t} \in \mathbb{C}^{ N_{\text{tx}}} \).
The noisy received signal at the \gls{mt}'s antennas is given by
\begin{equation*}
    \bm{y} = \bm{H}\bm{t}s + \bm{n} \quad \in \mathbb{C}^{N_{\text{rx}}} \,,
\end{equation*}
where \( \bm{n} \sim \mathcal{N}_{\mathbb{C}}(\bm{0}, \sigma_{\bm{n}}^2 \mathbf{I}) \) represents the \gls{awgn}.
After receiving the noisy signal, the \gls{mt} performs an equalization by multiplying the received signal \( \bm{y} \) with the combining vector \( \bm{g} \in \mathbb{C}^{N_{\text{rx}}} \), obtaining the estimated symbol
\begin{equation*}
    \hat{s} = \bm{g}^{\mathrm{H}}\bm{H}\bm{t}s+\bm{g}^{\mathrm{H}}\bm{n} \,.
\end{equation*}
The precoder \( \bm{t} \) and the combiner \( \bm{g} \) are subject to the power constraints \( \lVert \bm{t} \rVert^2 < \rho_{\text{tx}} \) and \( \lVert \bm{g} \rVert^2 < \rho_{\text{rx}} \), and are designed to maximize the \gls{snr} of the received signal \( \text{SNR} = \mathbb{E}[\lvert \hat{s} \rvert]^2/\sigma_{\bm{n}}^2 \).

Generally, the precoding and combining vectors are chosen from a predefined codebook. 
This codebook often consists of columns from \gls{dft} matrices.
In the case of downlink precoders, the codebook entries are the columns \( \bm{f}_i \) of the \gls{dft} matrix \( \bm{F}_{\text{tx}} = \begin{bmatrix} \bm{f}_{\text{tx},1} & \cdots & \bm{f}_{\text{tx},N_{\text{tx}}} \end{bmatrix}
\).
In ideal single-user noiseless conditions, after appliying the beam alignment procedure, the \gls{bs} will select the combiner leading to the highest received signal power given by the index
\begin{equation*}
    i^\star = \arg\max_{i=1,\ldots,N_{\text{tx}}} \lVert \bm{H}\bm{f}_{\text{tx},i} \rVert^2 \,.
\end{equation*}
In our work, we center our perspective on the \gls{mt}, therefore we assume that the \gls{bs} applies the best precoder \( \bm{f}_{\text{tx},i^\star} \), which leads to the \gls{mt} experiencing an equivalent \gls{simo} channel \( \bm{\tilde{h}} = \bm{H}\bm{f}_{\text{tx},i^\star} \).
This assumption is reasonable if the \gls{bs} performs a beam sweeping procedure, as commonly specified in \gls{5g}-\gls{nr}, while the \gls{mt} independently explores different combiners.
If we assume that \( s \) is a known pilot symbol with unit norm, the \gls{mt} can potentially retrieve a noisy channel realization, and we can therefore express our equivalent system model as
\begin{equation} \label{eq:equivalent-simo-observation}
    \bm{\tilde{y}} =  \bm{\tilde{h}} + \bm{n} \,.
\end{equation}
Please note that we can also assume optimal precoding at the \gls{bs}, which is available in analytical form for the single-stream point-to-point \gls{mimo} case, without loss of generality.

Given the equivalent system model, we can outline the combining task.
Similarly to the \gls{bs}, we assume that the \gls{mt} employs a codebook defined by the columns of the \gls{dft} matrix \( \bm{F}_{\text{rx}} = \begin{bmatrix} \bm{f}_{\text{rx},1} & \cdots & \bm{f}_{\text{rx},N_{\text{rx}}} \end{bmatrix} \).
Please note that this assumption is made without loss of generality, and other codebooks can be selected.
To increase the number of \gls{dft} codebook entries, which is limited in the case of a reduced number of antennas \( N_{\text{rx}} \), we also consider \( N \)-times oversampled \gls{dft} codebooks consisting of \( N_{\text{c}} = N \cdot N_{\text{rx}} \) entries.
It is worth noting that \glspl{mt} are usually equipped with low-complexity hardware, with a single \gls{rf} chain and passive phase shifters at the antenna ports, without gain amplifiers.
This restricts the codebook to combiners of the form \( \bm{g} = \begin{bmatrix} e^{\mathrm{j}\phi_1} & \cdots & e^{\mathrm{j}\phi_{N_{\text{rx}}}} \end{bmatrix} \).
The selected \gls{dft} codebooks satisfy such a constraint.

Within this preliminary work, we assume a fixed orientation of the \gls{mt} antenna arrays.
While this assumption may not fully capture the dynamics of handheld \glspl{mt} in cellular networks, it remains realistic for systems where \glspl{mt} are fixed on vehicles and robots moving along specified trajectories.

\section{Proposed Method}
    \subsection{Preliminaries on the Coupled GMM}
    The work in~\cite{noPilotsNoProblem-Weisser2025} proposed the usage of a coupled \gls{gmm}-based model to obtain a generative prior, which is leveraged for downlink multi-user precoding.
    The usage of a coupled \gls{gmm} is particularly handy because it provides access to the analytic expression of the joint distribution.
    In our case, we use the \glspl{mt}' channel observations \( \bm{\tilde{y}} \) defined in \eqref{eq:equivalent-simo-observation} together with its position \( \bm{r} \in \mathbb{R}^{N_{\text{r}}} \) to train a coupled \gls{gmm} model that approximates the joint distribution 
    \begin{equation} \label{eq:joint-distribution}
        p(\bm{\tilde{h}}, \bm{r}) = \sum_{k=1}^K \pi_k \, \mathcal{N}_{\mathbb{C}}(\bm{\tilde{h}}; \bm{0},\bm{C}_{\bm{\tilde{h}},k}) \, \mathcal{N}_{\mathbb{R}}(\bm{r}; \bm{\mu}_{\bm{r},k}, \bm{C}_{\bm{r},k})
    \end{equation}
    between the equivalent \gls{simo} channel \( \bm{\tilde{h}} \) and the position \( \bm{r} \).
    The hyperparameter \( K \) is the number of Gaussian mixtures.
    The variables \( \bm{C}_{\bm{\tilde{h}},k} \), \( \bm{\mu}_{\bm{r},k} \), and \( \bm{C}_{\bm{r},k} \), represent, respectively, the \gls{ccm}, the mean of the position, and the covariance of the position for the \( k \)-th \gls{gmm} component.
    The variable \( \pi_k \) represents the so-called \gls{gmm} mixing coefficients, and allows us to model the discrete categorical distribution \( p(k) \). 
    The variable \( k \) is a latent variable that allows defining the conditionally Gaussian distributions 
    \begin{align*}
        p(\bm{\tilde{h}} \mid k) & = \mathcal{N}_{\mathbb{C}}(\bm{\tilde{h}}; \bm{0},\bm{C}_{\bm{\tilde{h}},k})\,, \\
        p(\bm{r} \mid k) & = \mathcal{N}_{\mathbb{R}}(\bm{r}; \bm{\mu}_{\bm{r},k}, \bm{C}_{\bm{r},k}) \,.
    \end{align*}
    
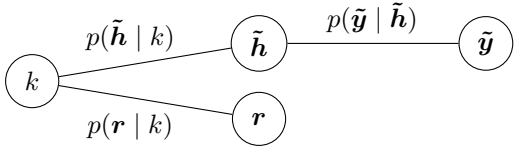
\begin{figure}
        \centering
        \begin{tikzpicture}
            \node[draw, circle, color=black, minimum size=20pt] (k) at (0,-0.5) {$k$};
            \node[draw, circle, color=black, minimum size=20pt] (h) at (3,0) {$\bm{\tilde{h}}$};
            \node[draw, circle, color=black, minimum size=20pt] (y) at (6,0) {$\bm{\tilde{y}}$};
            \node[draw, circle, color=black, minimum size=20pt] (r) at (3,-1) {$\bm{r}$};
    
            \node[] () at (1.3, 0.1) {$p(\bm{\tilde{h}}\mid k)$};
            \node[above] () at (4.5, 0) {$p(\bm{\tilde{y}}\mid \bm{\tilde{h}})$};
            \node[] () at (1.3, -1.1) {$p(\bm{r}\mid k)$};
            
            \draw[-] (k) -- (h);
            \draw[-] (k) -- (r);
            \draw[-] (h) -- (y);
        \end{tikzpicture}
        \caption{Probabilistic graphical model representing the assumed statistical structure, where the latent variable \( k \) induces conditional independence between \( \bm{\tilde{h}} \) and \( \bm{r} \), and \( \bm{\tilde{y}} \) is conditionally dependent on \( \bm{\tilde{h}} \).}
        \label{fig:probabilistic-graph}
    \end{figure}

    Please notice that the latent variable \( k \) is assumed to render the conditional distributions \( p(\bm{\tilde{h}} \mid k) \) and \( p(\bm{r} \mid k) \) independent from each other.
    This conditional independence structure is also evident from the probabilistic graph in \Cref{fig:probabilistic-graph}.
    An implicit assumption of the conditional independence modeling is that the second‑order statistics of the equivalent \gls{simo} channel vary slowly with the \gls{mt}’s position, such that the \gls{ccm} can be considered constant within a small spatial neighborhood.
    This follows from a local \gls{wss} property of the spatial wireless channel, which is justified by the environment scattering structure that changes negligibly over small displacements.
    Such property was exploited in~\cite{noPilotsNoProblem-Weisser2025} considering the wireless channel measured by the \gls{bs}; however, it can also be exploited at the \gls{mt}, with the only difference being a richer scattering environment near the \gls{mt} due to the vicinity to the scatterers and the related larger angular spread values~\cite[Table 7.5.6]{3gpp.38.901.1610}.

    Furthermore, considering \eqref{eq:equivalent-simo-observation}, we define the conditional distribution 
    \begin{equation*}
        p(\bm{\tilde{y}}\mid k) = \mathcal{N}_{\mathbb{C}}(\bm{\tilde{y}}; \bm{0}, \bm{C}_{\bm{\tilde{h}},k} + \sigma_{\bm{n}}^2 \mathbf{I}) \,,
    \end{equation*}
    which is consistent with the conditional dependence structure illustrated in \Cref{fig:probabilistic-graph}.
    Such conditional distribution allows us to analytically define the joint distribution \( p(\bm{\tilde{y}},\bm{r})\) by replacing \( p(\bm{\tilde{h}}\mid k) \) with \( p(\bm{\tilde{y}}\mid k) \) in \eqref{eq:joint-distribution}.

    \textit{Training Procedure}:
    In the proposed scheme we consider the availability of a dataset \( \mathcal{D} = \{ (\bm{\tilde{y}}_{\ell}, \bm{r}_{\ell}) \}_{\ell=1}^L \), consisting of samples that are represented by tuples of noisy channel measurements \( \bm{\tilde{y}}_{\ell} \) and measured positions \( \bm{r}_{\ell} \), obtained for example via \gls{gnss}.
    We follow the training of the coupled \gls{gmm} as defined in \cite{noPilotsNoProblem-Weisser2025}, where the model parameters \( \bm{\theta} = \{ \bm{C}_{\bm{\tilde{h}},k}, \bm{\mu}_{\bm{r},k}, \bm{C}_{\bm{r},k}, \pi_k \}_{k=1}^K \) are optimized via the maximization of the data log-likelihood, which, for latent variable models such as the coupled \gls{gmm} is done via the \gls{em} algorithm.
    The algorithm iterates between an expectation (E-step) and a maximization (M-step) that can be summarized as
    \begin{equation*}
        \bm{\theta}^{(t+1)} = \arg\max_{\bm{\theta} \in \bm{\Theta}} \mathbb{E}_{p(k \mid \bm{\tilde{h}},\bm{r}; \bm{\theta}^{(t)})} \left[ \log p( \bm{\tilde{h}},\bm{r}, k; \bm{\theta}) \right] \,,
    \end{equation*}
    where \( t \) represents the iteration index.
    The E-step computes the posterior probabilities of \( k \) for each sample given the parameters at iteration \( t \), which is given by the expression
    \begin{equation*}
        p(k \mid \bm{\tilde{y}}_{\ell}, \bm{r}_{\ell}; \bm{\theta}^{(t)}) = \frac{ \pi_k^{(t)} p(\bm{y}_{\ell} \mid k;\bm{\theta}^{(t)}) \, p(\bm{r}_{\ell} \mid k;\bm{\theta}^{(t)}) }{p(\bm{y}_{\ell}, \bm{r}_{\ell}; \bm{\theta}^{(t)})} \,,
    \end{equation*}
    where all the terms can be analytically expressed~\cite{noPilotsNoProblem-Weisser2025}.
    These posterior probabilities, also called responsibilities, allow us to update the model parameters during the M-step as
    \begin{align}
        \pi_k^{(t+1)} & = N_k^{(t)} / L \nonumber \\
        \bm{C}_{\bm{\tilde{h}},k}^{(t+1)} & = P_{\mathcal{S}_0^+}\left( \frac{1}{N_k^{(t)}}\sum_{\ell =1}^L \left( p(k \mid \bm{\tilde{y}}_{\ell}, \bm{r}_{\ell}; \bm{\theta}^{(t)})\,\bm{\tilde{y}}\bm{\tilde{y}}^{\mathrm{H}} \right) - \sigma_{\bm{n}}^2\mathbf{I} \right) \nonumber \\
        \bm{\mu}_{\bm{r},k}^{(t+1)} & = \frac{1}{N_k^{(t)}} \sum_{\ell=1}^L p(k \mid \bm{\tilde{y}}_{\ell}, \bm{r}_{\ell}; \bm{\theta}^{(t)}) \, \bm{r}_{\ell} \label{eq:m-step-r-means} \\
        \bm{C}_{\bm{r}, k}^{(t+1)} & = \label{eq:m-step-r-covariances} \\ & \hspace{-0.2cm} \frac{1}{N_k^{(t)}} \sum_{\ell=1}^L p(k \mid \bm{\tilde{y}}_{\ell}, \bm{r}_{\ell}; \bm{\theta}^{(t)}) \, (\bm{r}_{\ell} - \bm{\mu}_{\bm{r},k}^{(t+1)})(\bm{r}_{\ell} - \bm{\mu}_{\bm{r},k}^{(t+1)})^{\mathrm{T}} \nonumber
    \end{align}
    where \( N_k^{(t)} = \sum_{\ell=1}^L p(k \mid \bm{\tilde{y}}_{\ell}, \bm{r}_{\ell}; \bm{\theta}^{(t)}) \), and \( P_{\mathcal{S}_0^+}(\cdot) \) is the projection on positive semidefinite matrices.

\subsection{Model Refinement for Fixed Codebooks} \label{subsec:model-refinement-for-fixed-codebooks}
    The \gls{em} algorithm converges after \( T \) iterations, giving access to the approximation of the joint distribution \( p(\bm{\tilde{h}}, \bm{r}; \bm{\theta}^{(T)}) \).

    For the \( k \)-th of the \( K \) \gls{gmm} components, the combiner maximizing the received signal \gls{snr} is defined by the principal eigenvector \( \bm{v}_{k,1} \) of the corresponding \gls{ccm} \( \bm{C}_{\bm{\tilde{h}},k} \)~\cite{efficientUseFadingMIMOsystems-ivrlac2001}.
    However, as discussed in \Cref{sec:system-model}, there is a set of constraints on the \gls{mt}'s hardware that restricts us to the usage of a specific codebook of combiners.
    Therefore, we select the codebook entry \( \bm{f}_{\text{rx},j_k} \) according to its projection onto the principal eigenvector by finding
    \begin{equation*}
        j_k = \arg\max_{j\in \{1,\ldots,N_{\text{c}}\}} \lvert \bm{f}_{\text{rx},j}^{\mathrm{H}} \bm{v}_{k,1} \rvert \,.
    \end{equation*}

    Performing this projection introduces a mismatch between the statistical model \( p(\bm{\tilde{h}}, \bm{r}; \bm{\theta}^{(T)}) \) and the codebook-based representation.
    In the trained model, each latent \( k \) is simultaneously associated with both the marginals, \( p(\bm{\tilde{h}} \mid k; \bm{\theta}^{(T)}) \) and \( p(\bm{r} \mid k; \bm{\theta}^{(T)}) \), whose structure is determined by the \glspl{ccm} \( \bm{C}_{\bm{\tilde{h}},k}^{(T)} \) and the positional parameters \( \{ (\bm{\mu}_{\bm{r},k}^{(T)}, \bm{C}_{\bm{r},k}^{(T)})\}_{k=1}^{K} \), respectively.
    After projecting onto a reduced eigen-subspace and mapping to a codebook entry, the \glspl{ccm}' structure is no longer fully consistent with the representation used by the codebook.
    To address this issue, we refine the trained parameters by updating the model \glspl{ccm} as
    \begin{equation*}
        \bm{C}_{\bm{\tilde{h}},k}^{(T+1)} = \bm{f}_{\text{rx},j_k} \bm{f}_{\text{rx},j_k}^{\mathrm{H}}\,.
    \end{equation*}
    These updated \glspl{ccm} are then used to compute the posterior probabilities
    \begin{equation*}
        p(k \mid \bm{\tilde{h}}_{\ell}, \bm{r}_{\ell}; \bm{\theta}^{(T+1)}) = \frac{ \pi_k^{(T)} p(\bm{y}_{\ell} \mid k;\bm{\theta}^{(T+1)}) , p(\bm{r}_{\ell} \mid k;\bm{\theta}^{(T)}) }{p(\bm{y}_{\ell}, \bm{r}_{\ell}; \bm{\theta}^{(T)})}\,.
    \end{equation*}
    Finally, the consistency between the newly obtained marginal \( p(\bm{\tilde{h}} \mid k; \bm{\theta}^{(T+1)}) \) and \( p(\bm{r} \mid k; \bm{\theta}^{(T)}) \) is obtained by updating \( \bm{\mu}_{\bm{r},k}^{(T+1)} \) and \( \bm{C}_{\bm{r},k}^{(T+1)} \) via \eqref{eq:m-step-r-means} and \eqref{eq:m-step-r-covariances}.
    Importantly, if the model were trained from scratch under the constraint \( \bm{C}_{\bm{\tilde{h}},k} =  \bm{f}_{\text{rx},j_k} \bm{f}_{\text{rx},j_k}^{\mathrm{H}} \), one would need to define the values \( j_k \) in advance, namely, predefining how many \gls{gmm} components are assigned to each of the \( N_{\text{c}} \) codebook entries.
    The proposed refinement completely avoids this manual allocation.

    \subsection{Online Usage of the Model}
    
    Given the position \( \bm{r}_{\ell} \) of the \( \ell \)-th data sample, the coupled \gls{gmm} enables the identification of the most likely Gaussian component as
    \begin{align}
        k_{\ell} &  = \arg \max_k p(k \,|\,\bm{r}_{\ell}; \bm{\theta}^{(T)}) \nonumber \\ 
            & = \arg \max_k \frac{\pi_k^{(T)} \, \mathcal{N}_\mathbb{R}(\bm{r}_{\ell};\bm{\mu}_{\bm{r},k}^{(T)},\bm{C}_{\bm{r},k}^{(T)})}{\sum_{i=1}^K\mathcal{N}_\mathbb{R}(\bm{r}_{\ell};\bm{\mu}_{\bm{r},i}^{(T)},\bm{C}_{\bm{r},i}^{(T)})} \,, \label{eq:gaussian-component-given-position} 
    \end{align}
    which is in turn directly mapped to the codebook entry index \( j_{k_{\ell}} \).

    This mapping can be interpreted as a function \( f_{\bm{\theta}^{(T+1)}}(\bm{r}_{\ell}) \in \{ 1, \ldots, N_{\mathrm{c}} \} \), that is parameterized by the statistical model's parameters \( \bm{\theta}^{(T + 1)} \) and that can be evaluated at any position value.
    Such mapping defines the proposed \gls{bim}.

    \subsection{Computational and Memory Efficiency}
    Please notice that the evaluation of the density functions in \eqref{eq:gaussian-component-given-position} for a low-dimensional position vector \( \bm{r} \), i.e., \( N_{\text{r}} \in \{2, 3\} \), is computationally lightweight and is parallelizable for the \( K \) Gaussian components.
    Furthermore, the parameters required to solve \eqref{eq:gaussian-component-given-position} that need to be offloaded on the end terminal consist of \( K (1 + N_{\text{r}} + N_{\text{r}}(N_{\text{r}}+1)/2) \) floating point parameters, making it particularly lightweight in terms of memory constraints.
    Such properties render the used model particularly suitable for low-complexity hardware requirements.

\section{Simulation Results}
    \subsection{Datasets}
    We construct two datasets for the training of our models, one based on the standardized DeepMIMO Boston 5G dataset~\cite{deepmimo2019}, while the other is based on the \gls{quadriga} geometry-based stochastic channel simulator~\cite{quadrigaTechReport-Jaeckel2023}.
    
    For both datasets, we consider a \gls{bs} equipped with an \gls{ula} of 32 antennas, and \glspl{mt} equipped with an \gls{ula} of 4 antennas, both at a half-wavelength spacing.
    The central frequency is \SI{3.5}{\giga\hertz}, and we consider a static and narrowband scenario.
    We train our models with \( L = 10^5 \) samples and we use additional \( L_{\text{test}} = 5\cdot10^3 \) samples for testing.

    The DeepMIMO Boston 5G scenario considers, similarly to~\cite{noPilotsNoProblem-Weisser2025}, a rotation of the \gls{bs}'s and \gls{mt}'s \glspl{ula} of \SI{-45}{\degree} relative to the \( z \)-axis. 
    Both \gls{bs} and \glspl{mt} use omnidirectional antennas.

    The \gls{quadriga} dataset is constructed considering users disposed uniformly over a square area of \SI{50}{\meter} side length, where the \gls{bs} is positioned in the middle of the left side, at a height of \SI{25}{\meter}.
    The \gls{bs} employs a 3gpp-3d antenna array~\cite{3gpp.36.873} oriented towards the right, while \glspl{mt} utilize omnidirectional antennas.
    Both \gls{bs} and \glspl{mt} utilize vertical polarization only.
    We consider the 3GPP\_38.901\_UMa\_NLOS \gls{nr} model.
    
    We simulate \( L \) \gls{mimo} channels \( \bm{H}_{\ell} \in \mathbb{C}^{4 \times 32} \), where \( \ell \in \{ 1, \ldots, L \} \). 
    After normalizing each sample \( \ell \) with respect to its own path-loss coefficient, we process them as explained in \Cref{sec:system-model} to obtain the equivalent \(4\)-dimensional \gls{simo} channels \( \bm{\tilde{h}}_{\ell}\).
    We then normalize the data to obtain a unitary average power per antenna leading to an \gls{snr} definition as \( \text{SNR} = \mathrm{E} \left[ \lVert \bm{\tilde{h}} \rVert^2 \right] / \sigma_{\bm{n}}^2 = N_{\text{rx}} / \sigma_{\bm{n}}^2 \).
    
    \subsection{Baseline Methods}
    We compare our method against the following baseline approaches.
    
    We first consider a clustering-based fingerprinting approach, where we define a clustering of the users based on their position by ignoring their \glspl{csi}.
    The clustering is performed via a \gls{gmm} trained solely on the position information, and by assigning a user to a cluster associated with the Gaussian component with the highest responsibility~\cite[(9.13)]{patternRecognitionMachineLearning-bishop-2007}.
    Note that the clustering could also be performed by applying the K-means algorithm or a simple grid-based division of the space.
    We compute for each cluster the sample-based \gls{ccm} \( \bm{\hat{C}}_{\bm{\tilde{y}},k} = \frac{1}{N_k}\sum_{\ell=1}^{N_k} \bm{\tilde{y}}_{\ell}\bm{\tilde{y}}_{\ell}^{\mathrm{H}} \) and assign a cluster-based fingerprint that is given by the projection on the closest codebook entry of the principal eigenvector of \( \bm{\hat{C}}_{\bm{\tilde{y}},k} \).

    We also compare with exhaustive search, in which all \( N_{\mathrm{c}} \) codebook entries \( \bm{f}_{\mathrm{rx}, j} \), \( j\in\{1, \ldots, N_{\mathrm{c}} \} \), are probed for each noisy measurement \( \bm{\tilde{y}}_{\ell} \). 
    The codebook index is then selected as
    \begin{equation*}
        j_{\text{exh}_{\ell}} =
        \arg\max_{j\in\{1, \ldots, N_{\mathrm{c}}\}} \lvert \bm{f}_{\mathrm{rx}, j}^{\mathrm{H}}\bm{\tilde{y}}_{\ell} \rvert^2 \,,
    \end{equation*}
    i.e., based on the probe yielding the highest received signal power.

    The genie selection approach selects a codebook entry assuming a perfect knowledge of the \gls{csi} by solving
    \begin{equation*} \label{eq:genie-codebook-selection}
        j_{\text{genie}_{\ell}} = \arg\max_{j\in\{1, \ldots, N_{\mathrm{c}}\}} \lvert \bm{f}_{\mathrm{rx}, j}^{\mathrm{H}}\bm{\tilde{h}}_{\ell} \rvert^2 \,.
    \end{equation*}
    
    Finally, a random selection of codebook entries is also considered, where the selected index \( j_{\text{rnd}, \ell} \) is randomly chosen from the set \( \{ 1, \ldots, N_{\text{c}} \} \) for each test sample.
    
    \subsection{Evaluation Metrics}
    We evaluate the performance by computing the accuracy metric on the test dataset \( \mathcal{D}_{\text{test}} = \{ (\bm{\tilde{y}}_{\ell}, \bm{r}_{\ell} ) \}_{\ell=1}^{L_{\text{test}}} \).
    This metric considers the fraction of test samples where the codebook entry selected by a specific method---i.e., \( j_{k_{\ell}} \) for the proposed method, \( j_{\text{exh}_{\ell}} \) for the exhaustive search, and \( j_{\text{rnd}_{\ell}} \) for the random selection---is corresponding to the genie selection \( j_{\text{genie}_{\ell}} \).
    That is, we compute
    \begin{equation*}
        A_{(\cdot)} = \frac{1}{L_{\text{test}}} \sum_{\ell=1}^{L_{\text{test}}} 1_{\{ j_{(\cdot)_{\ell}} = j_{\text{genie}_{\ell}} \}} \,,
    \end{equation*}
    where \( (\cdot) \in \{ k,\text{exh}, \text{rnd} \}\), and where \( 1_{\{\cdot\}}\) is the indicator function that is equal to 1 if the condition inside the curly brackets is true and 0 otherwise.

    We also consider the \gls{mnse} that is defined as 
    \begin{equation*}
        \mathrm{MnSE} = \frac{1}{L_{\text{test}}}\sum_{\ell=1}^{L_{\text{test}}} \frac{R(\bm{\tilde{h}}_{\ell},\bm{f}_{\text{rx},j_{(\cdot)_\ell}})}{R(\bm{\tilde{h}}_{\ell},\bm{\tilde{h}}_{\ell}/\lVert \bm{\tilde{h}}_{\ell}\rVert)} \,,
    \end{equation*}
    where
    \begin{equation*}
        R(\bm{\tilde{h}}_{\ell}, \bm{v}) = \log_2 \left( 1 + \frac{\lvert \bm{v}^{\text{H}} \bm{\tilde{h}}_{\ell}  \rvert^2 }{\sigma_{\bm{n}}^2} \right)\,.
    \end{equation*}

    \subsection{Performance Evaluation}
\begin{figure}
        \centering
        \includegraphics[width=\linewidth]{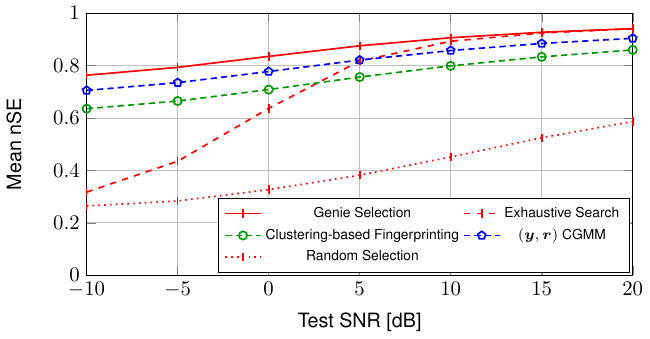}
        \vspace{-0.5cm}
        \caption{\gls{mnse} over \gls{snr}, DeepMIMO dataset, training \gls{snr} = \SI{10}{\deci\bel}, \gls{gmm} with \( K = 64 \) components and not oversampled 4-\gls{dft} codebook (\( N_{\text{c}} = 4 \)).} 
        \label{fig:mnse-vs-snr-deepmimo-train-10dB}
    \end{figure}

\begin{figure}
        \centering
        \includegraphics[width=\linewidth]{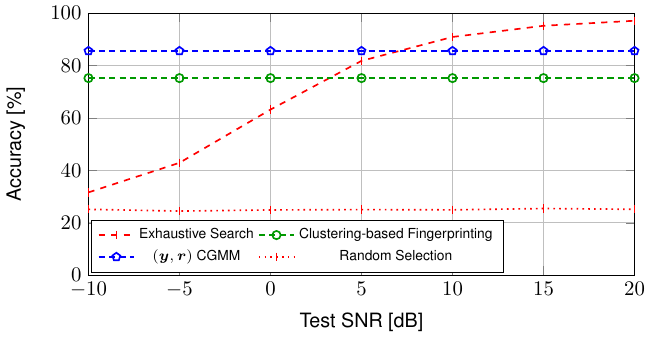}
        \vspace{-0.5cm}
        \caption{Accuracy over \gls{snr}, DeepMIMO dataset, training \gls{snr} = \SI{10}{\deci\bel}, \gls{gmm} with \( K = 64 \) components and not oversampled 4-\gls{dft} codebook (\( N_{\text{c}} = 4 \)).} 
        \label{fig:accuracy-vs-snr-deepmimo-train-10dB}
    \end{figure}

    \Cref{fig:mnse-vs-snr-deepmimo-train-10dB} and \Cref{fig:accuracy-vs-snr-deepmimo-train-10dB} represent the evaluation of the \gls{mnse} and the accuracy for the DeepMIMO dataset.
    It can be observed that the proposed method, denoted as ``\( (\bm{y}, \bm{r}) \) CGMM'', outperforms the clustering-based fingerprinting approach.
    This improvement arises because the proposed method incorporates \gls{csi} information into the statistical model, while the clustering-based fingerprinting approach disregards the statistical relation between the position and \gls{csi}.
    Exhaustive search performs better than the proposed method in the high-\gls{snr} regime, where high-quality channel observations enable an accurate selection of the codebook entry.
    However, the exhaustive search requires measuring the signal power for all codebook entries, leading to increased latency and energy consumption compared to the proposed method.
    
\begin{figure*}
        \centering

        \begin{minipage}{0.8\linewidth}
            \includegraphics[trim={0 6.5cm 0 1.8cm},clip,width=\linewidth]{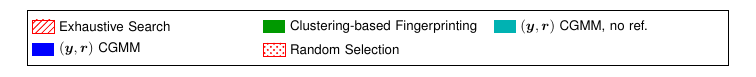}
        \end{minipage}
    
        \subfloat[DeepMIMO, training and test \glspl{snr} \SI{0}{\deci\bel}, \( K = 128 \), \\ \hspace*{0.35cm} 4 times oversampled 4-\gls{dft} codebook (\( N_{\text{c}} = 16 \)).]{
            \begin{minipage}{0.48\linewidth}
                \edef\SNR{0}
                \label{fig:selected-codebook-entry-deepmimo-0dB}
                \includegraphics[trim={0.5cm 2.7cm 0.5cm 2.3cm},clip,width=\linewidth]{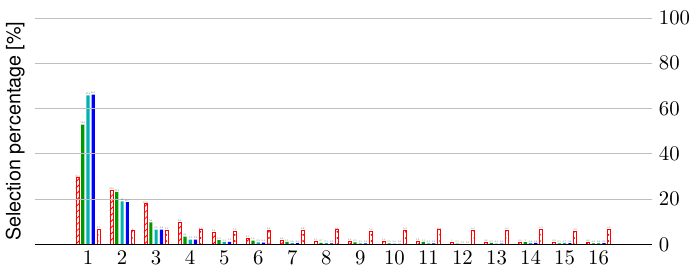}
            \end{minipage}
        }
        \subfloat[DeepMIMO, training and test \glspl{snr} \SI{10}{\deci\bel}, \( K = 128 \), \\ \hspace*{0.35cm} 4 times oversampled 4-\gls{dft} codebook (\( N_{\text{c}} = 16 \)).]{
            \begin{minipage}{0.48\linewidth}
                \edef\SNR{10}
                \label{fig:selected-codebook-entry-deepmimo-10dB}
                \includegraphics[trim={0.5cm 2.7cm 0.5cm 2.3cm},clip,width=\linewidth]{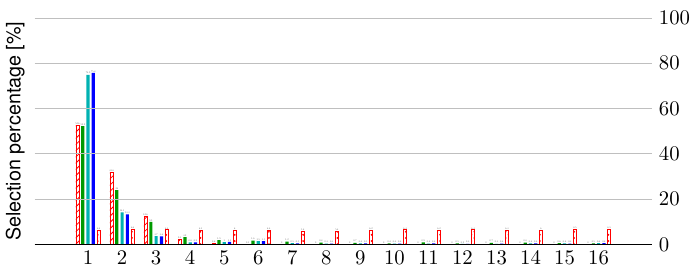}
            \end{minipage}
        }
        
        \subfloat[\gls{quadriga}, training and test \glspl{snr} \SI{0}{\deci\bel}, \( K = 32 \), \\
        \hspace*{0.35cm} not oversampled 4-\gls{dft} codebook (\( N_{\text{c}} = 4 \)).]{
            \begin{minipage}{0.48\linewidth}
                \edef\SNR{0}
                \label{fig:selected-codebook-entry-quadriga-0dB}
                \includegraphics[trim={1cm 2cm 1cm 3cm},clip,width=\linewidth]{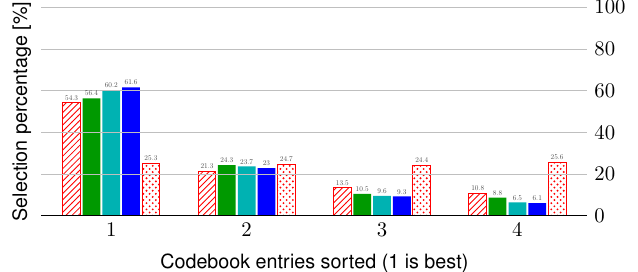}
            \end{minipage}
        }
        \subfloat[\gls{quadriga}, training and test \glspl{snr} \SI{10}{\deci\bel}, \( K = 32 \), \\
        \hspace*{0.35cm} not oversampled 4-\gls{dft} codebook (\( N_{\text{c}} = 4 \)).]{
            \begin{minipage}{0.48\linewidth}
                \edef\SNR{10}
                \label{fig:selected-codebook-entry-quadriga-10dB}
                \includegraphics[trim={1cm 2cm 1cm 3cm},clip,width=\linewidth]{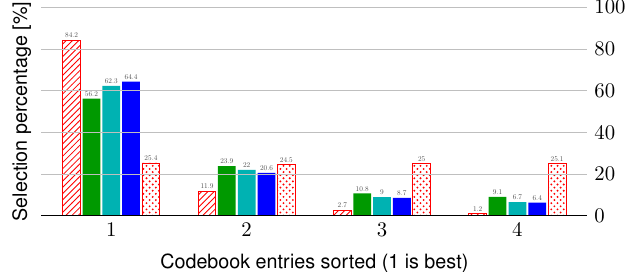}
            \end{minipage}
        }
        
        \caption{Codebook entry selected by each method.}
        \label{fig:codebook-entry-selected-by-each-method}
    \end{figure*}

    \Cref{fig:codebook-entry-selected-by-each-method} represents the selected codebook entries by each method for two \gls{snr} levels of the \gls{csi} observations, and for both datasets.
    In particular, the indices on the horizontal axis indicate the quality of the codebook entries, where index 1 indicates that the best codebook entry was selected and the last index the worst.
    As expected, the performance of the exhaustive search degrades at low \gls{snr} or with a high number of codebook entries.
    The proposed coupled \gls{gmm} method leads always to a performance increase compared to its variant without the refinement procedure described in \Cref{subsec:model-refinement-for-fixed-codebooks}, which is denoted as ``\( (\bm{y}, \bm{r}) \) CGMM, no ref.''
    In all cases, clustering-based fingerprinting tends to select poorer codebook entries with higher probability.

\begin{figure*}
        \centering

        \subfloat[Genie selection.]{
            \begin{minipage}{0.28\linewidth}
                \includegraphics[trim={0 0.6cm 4cm 1.8cm},clip,height=0.8\linewidth]{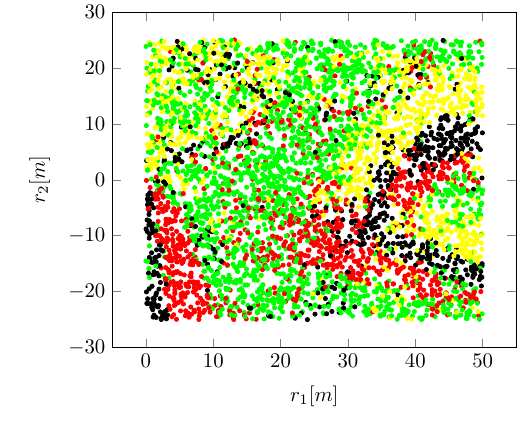}
                \label{fig:selected-codebook-entries-quadriga-10dB-genie}
            \end{minipage}
        }
        \subfloat[\( (\bm{y}, \bm{r}) \) CGMM, \( K = 64 \) components.]{
            \begin{minipage}{0.28\linewidth}
                \includegraphics[trim={0 0.6cm 4.7cm 1.8cm},clip,height=0.8\linewidth]{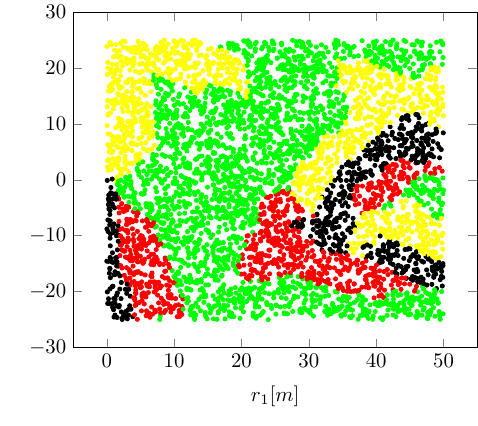}
                \label{fig:selected-codebook-entries-quadriga-10dB-y_r_cgmm}
            \end{minipage}
        }
        \subfloat[Clustering-based fingerprinting, 64 clusters.]{
            \begin{minipage}{0.28\linewidth}
                \includegraphics[trim={0 0.6cm 4.7cm 1.8cm},clip,height=0.8\linewidth]{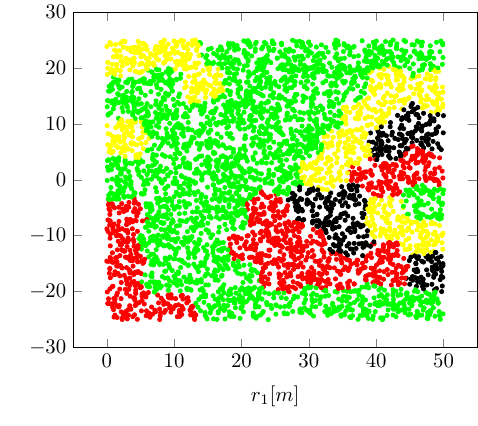}
                \label{fig:selected-codebook-entries-quadriga-10dB-clustering-based-fingerprinting}
            \end{minipage}
        }
        \caption{Selected codebook entries for each test sample position in the \gls{quadriga} dataset. Training and test \gls{snr} of \SI{10}{\deci\bel}.} \label{fig:selected-codebook-entries-quadriga-10dB}
    \end{figure*}

    \Cref{fig:selected-codebook-entries-quadriga-10dB} illustrates the selected codebook entry for each test sample position in the \gls{quadriga} dataset.
    \Cref{fig:selected-codebook-entries-quadriga-10dB-genie} shows the genie-aided selection, and highlights the complexity of the \gls{nlos} environment.
    Although certain regions favor specific codebook entries, position-based selection remains challenging due to small-scale fading, which can lead to different codebook choices even within areas where most users would select the same entry.
    \Cref{fig:selected-codebook-entries-quadriga-10dB-y_r_cgmm} presents the codebook selection obtained by the proposed \gls{bim}, which closely matches the genie-aided results, unlike the clustering-based fingerprinting shown in \Cref{fig:selected-codebook-entries-quadriga-10dB-clustering-based-fingerprinting}, which only gives a coarse matching.

\begin{figure*}
        \centering

        \begin{minipage}{0.95\linewidth}
            \includegraphics[trim={0 5.0cm 0 0.0cm},clip,width=\linewidth]{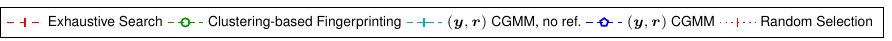}
        \end{minipage}
    
        \subfloat[DeepMIMO, training \gls{snr} and test \gls{snr} \SI{10}{\deci\bel}, \( K = 64 \), \\ \hspace*{0.35cm} 4 times oversampled 4-\gls{dft} codebook (\( N_{\text{c}} = 16 \)).]{
            \begin{minipage}{0.48\linewidth}
                \edef\SNR{0}
                \label{fig:accuracy-vs-sigma_r-deepmimo-10dB-k64-4oversampling}
                \includegraphics[trim={0.0cm 0.0cm -1.0cm 0.0cm},clip,width=\linewidth]{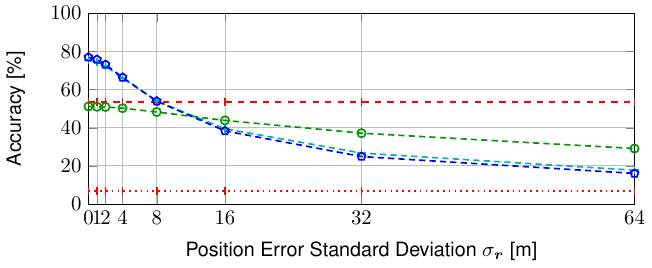}
            \end{minipage}
        }
        \subfloat[\gls{quadriga}, training and test \glspl{snr} \SI{10}{\deci\bel}, \( K = 64 \), \\ \hspace*{0.35cm} not oversampled 4-\gls{dft} codebook (\( N_{\text{c}} = 4 \)).]{
            \begin{minipage}{0.48\linewidth}
                \edef\SNR{10}
                \label{fig:accuracy-vs-sigma_r-quadriga-10dB-k64-no_oversampling}
                \includegraphics[trim={0.0cm 0.0cm -1.0cm 0.0cm},clip,width=\linewidth]{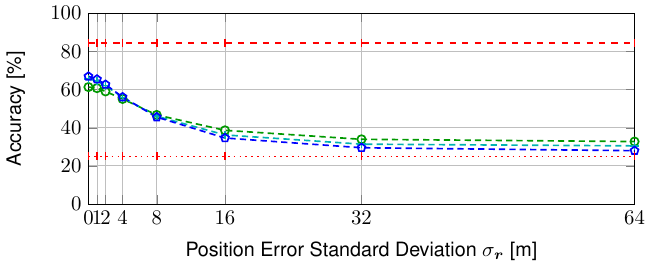}
            \end{minipage}
        }
        
        \caption{Accuracy values varying the error on the position information.}
        \label{fig:accuracy-vs-sigma_r}
    \end{figure*}

    \Cref{fig:accuracy-vs-sigma_r} presents the accuracy achieved by the compared methods when position estimation errors are introduced at inference time.
    The coupled \gls{gmm} approach benefits from low positioning errors and can outperform the clustering-based fingerprinting.
    However, the fingerprinting approach exhibits greater robustness as the positioning error increases. 
    In scenarios characterized by large positioning errors or rich scattering conditions, such as the \gls{quadriga} \gls{nlos} environment shown in \Cref{fig:accuracy-vs-sigma_r-quadriga-10dB-k64-no_oversampling}, the exhaustive search method becomes advantageous, but with the cost of higher computational complexity.
    It is worth noting that, in \Cref{fig:accuracy-vs-sigma_r-deepmimo-10dB-k64-4oversampling}, the exhaustive approach must probe \( N_{\text{c}} = 16 \) codebook entries, whereas all other methods can directly select the most suitable codebook entry, resulting in a substantial complexity reduction.
    
\section{Conclusion}
    We presented a position-aided beam selection method based on a coupled \gls{gmm}, enabling the construction of a \gls{bim} that can be executed directly at the \gls{mt} to efficiently perform combiner selection.
    The model is constructed by modeling the joint probability distribution of the \gls{csi} and the position of users.
    A refinement step was introduced to ensure consistency between the learned statistical model and the adopted codebook.
    The proposed solution is extremely lightweight in terms of memory and computational complexity, making it suitable for deployment on resource-constrained devices.
    Numerical results show that the proposed solution provides an alternative to extensive search approaches and outperforms clustering-based fingerprinting approaches.

    In future works, we want to modify the proposed model to account for the \glspl{mt}'s orientation by adjusting the corresponding \glspl{ccm} of the proposed statistical model.

\bibliographystyle{IEEEtran}

\bibliography{mybib}

@ARTICLE{VisionOf6GWirelessSystems-saad2020,
    author={Saad, Walid and Bennis, Mehdi and Chen, Mingzhe},
    journal={IEEE Network}, 
    title={A Vision of {6G} Wireless Systems: Applications, Trends, Technologies, and Open Research Problems}, 
    year={2020},
    volume={34},
    number={3},
    pages={134-142},
    doi={10.1109/MNET.001.1900287}
}

@ARTICLE{mmWaveBeamforming-Hur2013,
  author={Hur, Sooyoung and Kim, Taejoon and Love, David J. and Krogmeier, James V. and Thomas, Timothy A. and Ghosh, Amitava},
  journal={IEEE Transactions on Communications}, 
  title={Millimeter Wave Beamforming for Wireless Backhaul and Access in Small Cell Networks}, 
  year={2013},
  volume={61},
  number={10},
  pages={4391-4403},
  doi={10.1109/TCOMM.2013.090513.120848}}

@ARTICLE{6gWirelessNetworks-zhang2019,
  author={Zhang, Zhengquan and Xiao, Yue and Ma, Zheng and Xiao, Ming and Ding, Zhiguo and Lei, Xianfu and Karagiannidis, George K. and Fan, Pingzhi},
  journal={IEEE Vehicular Technology Magazine}, 
  title={{6G} Wireless Networks: Vision, Requirements, Architecture, and Key Technologies}, 
  year={2019},
  volume={14},
  number={3},
  pages={28-41},
  doi={10.1109/MVT.2019.2921208}}

@ARTICLE{TerahertzCommunicationsFor6gSurvey-Jiang2024,
  author={Jiang, Wei and Zhou, Qiuheng and He, Jiguang and Habibi, Mohammad Asif and Melnyk, Sergiy and El-Absi, Mohammed and Han, Bin and Renzo, Marco Di and Schotten, Hans Dieter and Luo, Fa-Long and El-Bawab, Tarek S. and Juntti, Markku and Debbah, Mérouane and Leung, Victor C. M.},
  journal={IEEE Communications Surveys \& Tutorials}, 
  title={Terahertz Communications and Sensing for {6G} and Beyond: A Comprehensive Review}, 
  year={2024},
  volume={26},
  number={4},
  pages={2326-2381},
  doi={10.1109/COMST.2024.3385908}}

@ARTICLE{IEEE802.11ad-2012,
  author={},
  journal={IEEE Std 802.11ad-2012 (Amendment to IEEE Std 802.11-2012, as amended by IEEE Std 802.11ae-2012 and IEEE Std 802.11aa-2012)}, 
  title={{IEEE} Standard for Information technology--Telecommunications and information exchange between systems--Local and metropolitan area networks--Specific requirements-Part 11: Wireless {LAN} Medium Access Control ({MAC}) and Physical Layer ({PHY}) Specifications Amendment 3: Enhancements for Very High Throughput in the 60 {GHz} Band}, 
  year={2012},
  volume={},
  number={},
  pages={1-628},
  doi={10.1109/IEEESTD.2012.6392842}}

@ARTICLE{IEEE802.11ay-2021,
  author={},
  journal={IEEE Std 802.11ay-2021 (Amendment to IEEE Std 802.11-2020 as amendment by IEEE Std 802.11ax-2021)}, 
  title={{IEEE} Standard for Information Technology--Telecommunications and Information Exchange between Systems Local and Metropolitan Area Networks--Specific Requirements Part 11: Wireless {LAN} Medium Access Control ({MAC}) and Physical Layer ({PHY}) Specifications Amendment 2: Enhanced Throughput for Operation in License-exempt Bands above 45 {GHz}}, 
  year={2021},
  volume={},
  number={},
  pages={1-768},
  doi={10.1109/IEEESTD.2021.9502046}}

@techreport{3gpp.38.843rel19,
    author = {{3GPP}},
    title = {Study on Artificial Intelligence ({AI})/Machine Learning ({ML}) for {NR} air interface},
    institution = {3rd Generation Partnership Project (3GPP)},
    series = {TR},
    number = {38.843},
    version = {19.0.0},
    year = {2025},
    month = {September},
    type = {Technical Report},
    url = {https://www.3gpp.org/dynareport/38843.htm}
}

@INPROCEEDINGS{beamTracking-Va2016,
  author={Va, Vutha and Vikalo, Haris and Heath, Robert W.},
  booktitle={2016 IEEE Global Conference on Signal and Information Processing (GlobalSIP)}, 
  title={Beam tracking for mobile millimeter wave communication systems}, 
  year={2016},
  volume={},
  number={},
  pages={743-747},
  doi={10.1109/GlobalSIP.2016.7905941}}

@ARTICLE{robustBeamTracking-Jayaprakasam2017,
  author={Jayaprakasam, Suhanya and Ma, Xiaoxue and Choi, Jun Won and Kim, Sunwoo},
  journal={IEEE Communications Letters}, 
  title={Robust Beam-Tracking for mmWave Mobile Communications}, 
  year={2017},
  volume={21},
  number={12},
  pages={2654-2657},
  doi={10.1109/LCOMM.2017.2748938}}

@INPROCEEDINGS{noPilotsNoProblem-Weisser2025,
  author={Weißer, Franz and Kasibovic, Amar and Böck, Benedikt and Utschick, Wolfgang},
  booktitle={2025 28th International Workshop on Smart Antennas (WSA)}, 
  title={No Pilots, No Problem: A Generative Model for Position-Based Downlink Precoding}, 
  year={2025},
  volume={},
  number={},
  pages={127-132},
  doi={10.1109/WSA65299.2025.11202852}}

@INPROCEEDINGS{probabilisticPositionAidedBeamSelection-Chege2025,
  author={Chege, Joseph K. and Yeredor, Arie and Haardt, Martin},
  booktitle={2025 33rd European Signal Processing Conference (EUSIPCO)}, 
  title={Probabilistic Position-Aided Beam Selection for {mmWave} {MIMO} Systems}, 
  year={2025},
  volume={},
  number={},
  pages={2107-2111},
  doi={10.23919/EUSIPCO63237.2025.11226390}}

@ARTICLE{environmentAwareChannelKnowledgeMap-Zeng2024,
  author={Zeng, Yong and Chen, Junting and Xu, Jie and Wu, Di and Xu, Xiaoli and Jin, Shi and Gao, Xiqi and Gesbert, David and Cui, Shuguang and Zhang, Rui},
  journal={IEEE Communications Surveys \& Tutorials}, 
  title={A Tutorial on Environment-Aware Communications via Channel Knowledge Map for {6G}}, 
  year={2024},
  volume={26},
  number={3},
  pages={1478-1519},
  doi={10.1109/COMST.2024.3364508}}

@INPROCEEDINGS{environmentAwareBeamAlignment-Wu2021,
  author={Wu, Di and Zeng, Yong and Jin, Shi and Zhang, Rui},
  booktitle={2021 IEEE International Conference on Communications Workshops (ICC Workshops)}, 
  title={Environment-Aware and Training-Free Beam Alignment for mmWave Massive {MIMO} via Channel Knowledge Map}, 
  year={2021},
  volume={},
  number={},
  pages={1-7},
  doi={10.1109/ICCWorkshops50388.2021.9473871}}

@techreport{quadrigaTechReport-Jaeckel2023,
    author={Jaeckel, Stephan and others},
    institution={{Fraunhofer Heinrich Hertz Institute}},
    note={{V. 2.8.1}},
    title={{QuaDRiGa} - Quasi Deterministic Radio Channel
    Generator, User Manual and Documentation},
    type={Technical Report},
    year={2023}
}

@techreport{3gpp.38.901.1610,
    author = {{3GPP}},
    title = {Study on channel model for frequencies from 0.5 to 100 {GHz}},
    institution = {3rd Generation Partnership Project (3GPP)},
    type = {Technical Report},
    number = {TR 38.901},
    version = {18.1.0},
    release = {18},
    year = {2026},
    month = {January},
    url = {https://www.3gpp.org/dynareport/38901.htm}
}

@techreport{3gpp.36.873,
    author = {{3GPP}},
    title = {Study on {3D} channel model for LTE},
    institution = {3rd Generation Partnership Project (3GPP)},
    type = {Technical Report},
    number = {TR 36.873},
    version = {12.5.0},
    release = {12},
    year = {2017},
    month = {June},
    url = {https://www.3gpp.org/dynareport/36873.htm}
}

@INPROCEEDINGS{efficientUseFadingMIMOsystems-ivrlac2001,
  author={Ivrlac, M.T. and Kurpjuhn, T.P. and Brunner, C. and Utschick, W.},
  booktitle={IEEE 54th Vehicular Technology Conference. VTC Fall 2001. Proceedings (Cat. No.01CH37211)}, 
  title={Efficient use of fading correlations in MIMO systems}, 
  year={2001},
  volume={4},
  number={},
  pages={2763-2767 vol.4},
  doi={10.1109/VTC.2001.957264}}

@InProceedings{deepmimo2019,
    author = {Alkhateeb, A.},
    title = {{DeepMIMO}: A Generic Deep Learning Dataset for Millimeter Wave and Massive {MIMO} Applications},
    booktitle = {Proc. of Information Theory and Applications Workshop (ITA)},
    year = {2019},
    pages = {1-8},
    month = {Feb},
    Address = {San Diego, CA}, }

@BOOK{patternRecognitionMachineLearning-bishop-2007,
      author       = {Bishop, Christopher M.},
      title        = {{P}attern {R}ecognition and {M}achine {L}earning},
      address      = {New York, NY},
      publisher    = {Springer Science+Business Media, LLC},
      reportid     = {ERESRWTH-2021-00110},
      isbn         = {978-0-387-31073-2},
      series       = {Information Science and Statistics},
      pages        = {Kapitel 9: S. 423-455},
      year         = {2006},
      cin          = {21ws-12.59324},
      ddc          = {006.4},
      cid          = {I:(DE-RON)21ws-12.59324},
      typ          = {PUB:(DE-HGF)3},
      doi          = {10.1007/978-0-387-45528-0},
}

@INPROCEEDINGS{multimodalDeepLearninMMWaveBeamPrediction-Shi2024,
  author={Shi, Binpu and Li, Min and Zhao, Ming-Min and Lei, Ming and Li, Liyan},
  booktitle={2024 IEEE 99th Vehicular Technology Conference (VTC2024-Spring)}, 
  title={Multimodal Deep Learning Empowered Millimeter-Wave Beam Prediction}, 
  year={2024},
  volume={},
  number={},
  pages={1-6},
  doi={10.1109/VTC2024-Spring62846.2024.10683225}}

\end{document}